\documentclass{elsart}
\usepackage{graphicx}

\usepackage{amssymb}

\begin{document}

\begin{frontmatter}
\title{%
    Single--particle spectrum for a model of fermions
    interacting with two-level local excitations on a lattice:
    A dynamical CPA approach
}

\author{%
    A.O.~Anokhin, A.V.~Zarubin and V.Yu.~Irkhin
}

\address{Institute of Metal Physics,
		Kovalevskaya str.18, 620041 Ekaterinburg, Russia}

\begin{abstract}
The problem of motion of a single electron interacting with a
periodic lattice of two-level systems is investigated within a
spinless fermion model. The Green's function is calculated in a
single-site dynamical coherent potential approximation
which is equivalent to
DMFT. The picture of one-electron density of states is obtained
for various values of the tunnel splitting and coupling between
the electron and two-level system. The occurrence of a band
splitting with increase of the coupling is demonstrated.
\end{abstract}

\end{frontmatter}

The problem of interaction of current carriers with local
excitations is one of the classical problem of solid state physics
\cite{Mahan:2000}. Recently, the interest in this topic has been
revived in connection with investigation of highly correlated
electron systems, in particular high-$T_{\mathrm{c}}$
superconductors where anharmonic lattice vibrations (i.e., those
of apical oxygen) are assumed sometimes to play an important role
\cite{Maruyama:1989:EXAFS_HTC,%
Mustre:1990:EXAFS_HTC,%
Mustre:1991:EXAFS_HTC,%
Irkhin:1991:PKLSA,%
Irkhin:1994:IKT}.
The
problem is relevant for both metals and semiconductors where the
electron spectrum is essentially modified by  the influence of
local excitations.

Formally the local excitations can be described in terms of
multi-level spectrum of a strongly anharmonic system. In the
simplest case one can use a model of two-level system (TLS) which
is formally described by pseudospin formalism. This model is
widely used for two-level systems in metallic glasses
\cite{Kondo:1976:MG,%
Zawadovsky:1980:MG,%
Keiter:1980:MG,%
Keiter:1984:MG},
crystalline-electric field (CEF) excitations
\cite{Fulde:1985:CEF},
dissipative systems with tunneling states
\cite{Leggett:1987:TLSD}.

In the present Letter we investigate the one-electron spectrum in
this model by using well established method --- single-site
dynamical  coherent potential approximation (DCPA)
\cite{Kakehashi:1992:DCPA1,%
Kakehashi:1992:DCPA2,%
Kakehashi:2002:DCPA3,%
Kakehashi:2004:CPAP} which becomes equivalent  to dynamical
mean-field theory (DMFT) \cite{Georges:1996:DMFT} in the limit of 
infinite dimensionality of the lattice 
($d\to\infty$, which
corresponds to Gaussian bare density of states
for hypercubic lattice)
or infinite
nearest neighbor number ($z\to\infty$, Bethe lattice). In those
limits the equivalence of both methods was demonstrated explicitly
in \cite{Kakehashi:2002:COMP}. From the practical point of view,
DCPA differs from DMFT by that the former retains the shape of the
bare density of states for finite $d$. Both the methods enable us
to map a rather complicated TLS lattice problem onto a
single-impurity problem which is readily solvable for one electron
in  empty conduction band.

We consider the spinless fermion model describing
fermion-pseudospin interaction:
\begin{equation}
H =
    \sum_{ij} t_{ij} c_{i}^{\dag} c_{j}
    - \sum_{i} (\mbox{\boldmath{$\gamma$}}_{i}
            \mbox{\boldmath{$\tau$}}_{i}) n_{i}
    - \sum_{i} (\mathbf{h}_{i}\mbox{\boldmath{$\tau$}}_{i})
\label{eq:H0}
\end{equation}
where    $c_{i}^{\dag}$, $c_{i}$ are fermion creation and
annihilation
    operators on a site $i$;
    $t_{ij}$ are transfer integrals,
    $n_{i}=c_{i}^{\dag}c_{i}$,
    $\mbox{\boldmath{$\tau$}}_{i}$ are pseudospin-$1/2$
    operators;
    $\mbox{\boldmath{$\gamma$}}_{i}$ is the vector of coupling between
    fermions and pseudospins, $\mathbf{h}_{i}$ is pseudomagnetic field vector,
the tunneling frequency being
$\Omega_{i}=\Delta_{i}=\sqrt{(\mathbf{h}_{i}\mathbf{h}_{i})}$.

Further on we consider translationally invariant case where site
dependence of both coupling constants and tunnelling frequency is
absent
\begin{equation}
H =
    \sum_{\mathbf{k}}
    	\epsilon_{\mathbf{k}} c_{\mathbf{k}}^{\dag} c_{\mathbf{k}}
    - \lambda \sum_{i} (\tau_{i}^{+}+\tau_{i}^{-}) n_{i}
    - h_{z} \sum_{i} \tau_{i}^{z} ,
\label{eq:H_k}
\end{equation}
with $\epsilon_{\mathbf{k}}$ being electron band spectrum, and we have
put $\gamma_{z}=0$ after pseudospin quantization axis rotation.

Note that the model (\ref{eq:H0},\ref{eq:H_k}) can be mapped onto
various models in different regions of its parameter space. For
example, at $\mbox{\boldmath{$\gamma$}}_{i}=(0,0,\gamma_z)$ and
$\mathbf{h}_{i}=(0,0,h_{z})$ we get the Falicov-Kimball model
\cite{Falikov:1969:SMSMT}. For large pseudospin values we obtain
by using the Holstein-Primakoff representation the Holstein model
\cite{Holstein:1959:MOD1,Holstein:1959:MOD2} which is widely used
to describe small polaron formation, in particular in molecular
crystals.

We calculate the one-particle Green's function in the case of
single conduction electron,
\begin{equation}
G_{\mathbf{k}}(z) =
    \langle\langle c_{\mathbf{k}}\vert
    c_{\mathbf{k}}^{\dag}\rangle\rangle_{z} =
        [z - \epsilon_{\mathbf{k}} - \Sigma_{\mathbf{k}}(z)]^{-1}
\label{eq:GF_k}
\end{equation}
where $\Sigma_{\mathbf{k}}(z)$  is the electron self-energy.

In the local self-consistent approximation (which corresponds to
dynamical CPA and  DMFT) the quantity $\Sigma_{\mathbf{k}}(z)$ is
replaced by the momentum-independent local self-energy,
$\Sigma_{\mathbf{k}}(z)\rightarrow\Sigma_{\mathrm{loc}}(z)$ and the
expression for the on-site Green's function
$G(z)=\sum_{\mathbf{k}}G_{\mathbf{k}}(z)$ takes the form
\begin{equation}
G(z) = R_{0}(z - \Sigma_{\mathrm{loc}}(z)), \quad R_{0}(z) =
    \sum_{\mathbf{k}} \frac{1}{z - \epsilon_{\mathbf{k}}} .
\label{eq:GF_appr}
\end{equation}
The local self-energy is obtained from the solution of the
auxiliary single-impurity problem, the corresponding model
parameters being determined from the self-consistency condition
$G(z) = G_{\mathrm{loc}}(z)$. The local Green's function is given by
\begin{equation}
G^{-1}_{\mathrm{loc}}(z) =
	\left[R_{0}^{\mathrm{loc}}(z)\right]^{-1} -\Sigma_{\mathrm{loc}}(z)
\label{eq:GF_locinv}
\end{equation}
where $R_{0}^{\mathrm{loc}}(z)$ is the  resolvent of the single-impurity
problem, with the interaction at the impurity site being switched
off. We obtain the self-consistency condition as
\begin{equation}
R_{0}^{\mathrm{loc}}(z) =
    \frac{G(z)}{1+G(z)\Sigma_{\mathrm{loc}}(z)} .
\label{eq:SCC}
\end{equation}
Introducing $F(z)$ by $F(R_{0}(z))=z$ to exclude $\Sigma_{\mathrm{loc}}(z)$
from the expression (\ref{eq:SCC}), one can write down the self-consistency
condition in a more familiar form
\begin{equation}
[R_{0}^{\mathrm{loc}}(z)]^{-1} = z - F(G(z)) + G^{-1}(z) ,
\label{eq:SCC_DMFT}
\end{equation}
if one prefers a DMFT-like formulation (see \cite{Georges:1996:DMFT}).

Further on we solve the auxiliary single-impurity problem with
the Hamiltonian on the fictitious one-dimensional lattice
\begin{eqnarray}
H  & =  &
	\sum_{n=0}^{\infty}
		\epsilon_{n} c_{n}^{\dag} c_{n}  
	+\sum_{n=0}^{\infty}
		\epsilon_{n,n+1} [c_{n}^{\dag} c_{n+1} + \mathrm{h.c.} ]  \nonumber \\
& &
	-\lambda c_{0}^{\dag}c_{0} (\tau^{+} + \tau^{-})
		-h_{z}\tau^{z}
\label{eq:H_imp}
\end{eqnarray}
by using the method described in
\cite{Cini:1988:ESEPRM}
to find $G_{\mathrm{loc}}(z)$ for a
single current carrier. Here $\epsilon_{n}$ and $\epsilon_{n,n+1}$ are
on-site energy levels and transfer integrals, respectively, for that fictitious
lattice, $n$ enumerates the lattice sites.
Then we write down the on-site impurity
scattering matrix for the pseudospin projection $\alpha$,
\begin{equation}
t_{\alpha}(z) =
\frac{
		\lambda^{2}R^{\mathrm{loc}}_{0}(z-\alpha\Delta)
	}{
		1-\lambda^{2}R^{\mathrm{loc}}_{0}(z-\alpha\Delta)
		R^{\mathrm{loc}}_{0}(z)
	} ,
\end{equation}
and introduce the average $T$-matrix
\begin{eqnarray}
t(z) & = & \sum_{\alpha=\pm} P_{\alpha} t_{\alpha}(z),   \\
P_{\sigma} &  = & \frac{1}{2} [ 1+\sigma\tanh(\beta\Delta/2)],
\quad
\Delta=h_{z},
\quad
\beta=\frac{1}{T} \nonumber
\end{eqnarray}
to obtain the standard expression
\begin{equation}
\Sigma_{\mathrm{loc}}(z) =
    \frac{t(z)}{1+R^{\mathrm{loc}}_{0}(z) t(z)} .
\label{eq:Sigma_t}
\end{equation}

As follows from the above treatment, $G_{\mathrm{loc}}(z)$ can be
considered as a solution to the problem of scattering by the
random substitutional impurity in the lattice with the bare
resolvent $R_{\mathrm{loc}}(z)$. In such a case, the quantity
$\lambda^{2}R^{\mathrm{loc}}_{0}(z-\alpha\Delta)$ has the meaning of the
random dynamical scattering potential for the impurity
distribution $P_{\alpha}$.

For the lattice model the quantity $\Sigma_{\mathrm{loc}}(z)$ calculated
under the assumption $R_{\mathrm{loc}}(z)=R_{0}(z)$ corresponds to the
average single-site T-matrix approximation (ATA), and the fully
self-consistent solution corresponds to CPA for the disordered
alloy problem \cite{Eliot:1974:EKL}. This analogy enables us to
use bonding and anti-bonding state classification.

We use the semielliptic bare density of states (the Bethe lattice
with infinite nearest-neighbor number which corresponds to DMFT
situation),
\begin{equation}
N(\epsilon)  =
    \frac{2}{\pi D^{2}}\sqrt{D^{2} - \epsilon^{2}},
\label{eq:SEDOS}
\end{equation}
$D$ being bare half-bandwidth,
and organize our numerical procedure as follows. For a given
$R_{0}^{\mathrm{loc}}(z)$ we calculate $\Sigma_{\mathrm{loc}}(z)$ from
(\ref{eq:Sigma_t}), and then $G(z)$ from
(\ref{eq:GF_k},\ref{eq:GF_appr}). Using the self-consistency
condition in the form (\ref{eq:SCC}) or (\ref{eq:SCC_DMFT})
we recompute updated $R_{0}^{\mathrm{loc}}(z)$ for
the next step of the iteration procedure. The iteration process
rapidly converges in a few steps. We also calculate $G_{\mathrm{loc}}(z)$
from (\ref{eq:GF_locinv}), but for the first iteration only.

As an initial condition for the iteration process we choose
$R_{0}^{\mathrm{loc}}(z)=R_{0}(z)$.
In that situation $G_{\mathrm{loc}}(z)$ obtained
at the first  iteration  gives us an exact solution of the
single-impurity problem on a lattice with the bare resolvent
$R_{0}(z)$ (i.e., the electron-pseudospin interaction is switched
off on all the sites but one). As for the first iteration lattice
$G(z)$, it has clear meaning of that calculated within the average
single-site $T$-matrix approximation, speaking in an alloy analogy
language.

Basing on the above consideration, we discuss the following three
approximations (a) single impurity site in the lattice (b)
dynamical ATA and (c) dynamical CPA.

\begin{figure}[tbp]
\centerline{%
\includegraphics[width=0.70\textwidth]{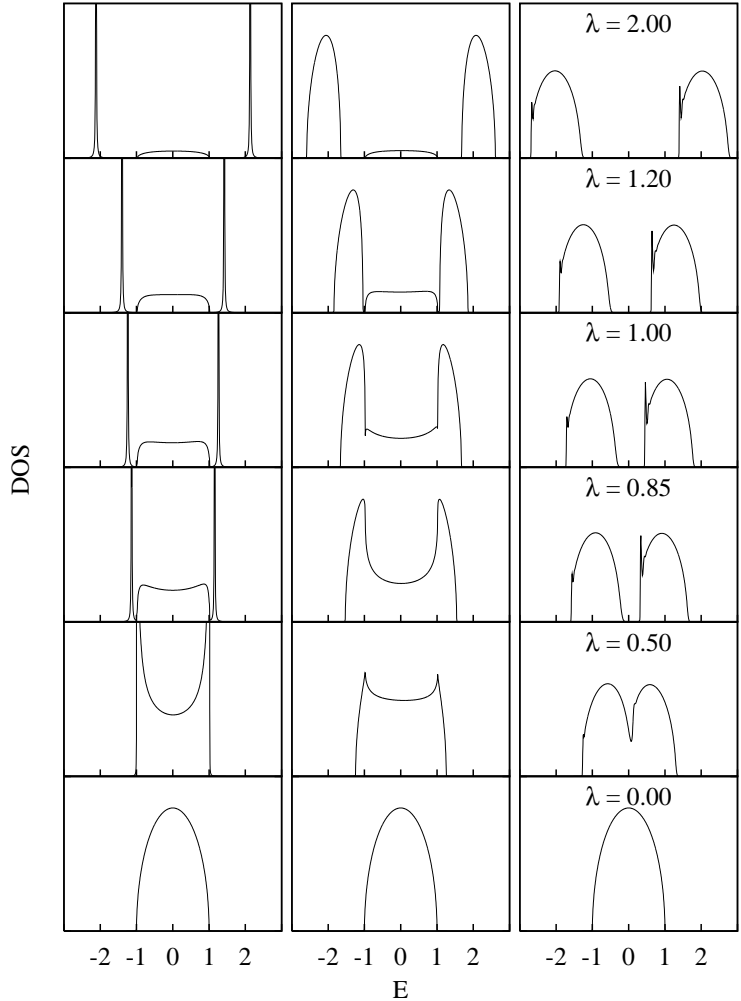}
}
\caption{%
Density of states for the semielliptic bare band with
$D=1$,  pseudomagnetic field $\Delta=0.02$ and different values of
coupling parameter $\lambda$ in three approximations (a) single
impurity in a lattice, left
(b) dynamical ATA, center
(c) dynamical CPA, right
}
\label{fig:1}
\end{figure}
\begin{figure}
\centerline{%
\includegraphics[width=0.70\textwidth]{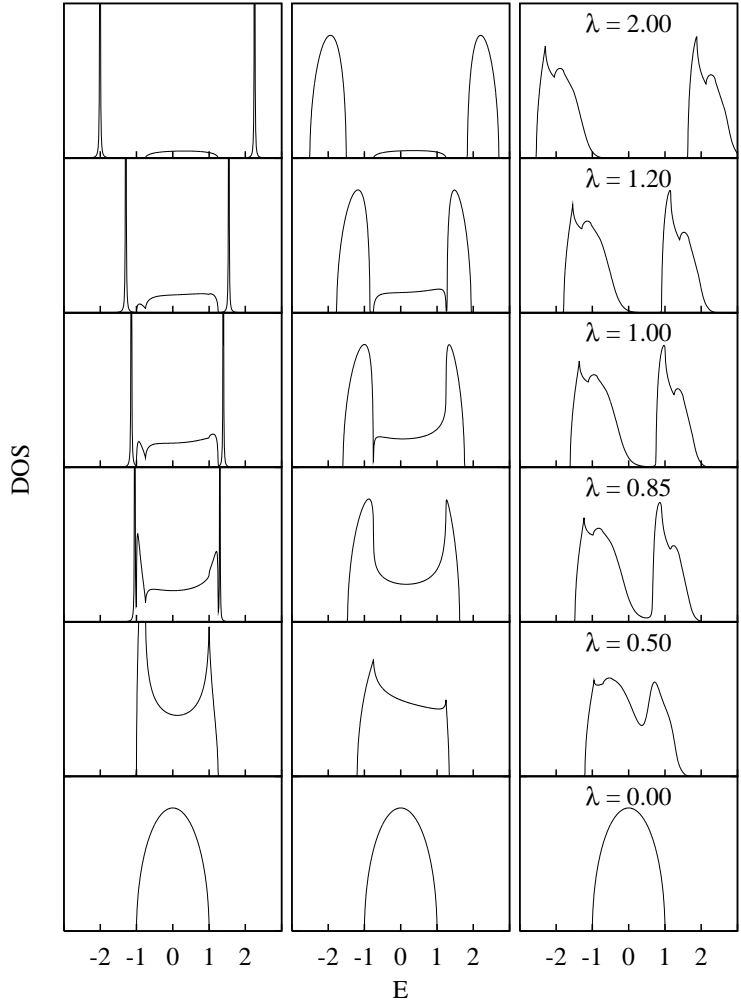}
}
\caption{%
The same data as in Fig.1 for $\Delta=0.25$
}
\label{fig:2}
\end{figure}
The results of calculations for zero temperature
are shown in Figs.\ref{fig:1}-\ref{fig:2}.
One can see that with
increasing $\lambda$ formation of bonding and anti-bonding states
takes place, which is clearly marked in the cases (a) and (b). It
should be noted that, unlike the scattering by static impurity
potential (where bonding and anti-bonding states occur depending
on the potential sign), dynamical nature of scattering in our
model results in coexistence of both the state types.

In the case (a) and for $0<\Delta<D$,
the critical values $\lambda_{\mathrm{c,b}}$ and
$\lambda_{\mathrm{c,a}}$ of the coupling parameter $\lambda$ for the
occurrence of bonding and anti-bonding states are given by
\begin{eqnarray}
\lambda_{\mathrm{c,b}} & = &
	\left[R_{0}(E_{\mathrm{b}})R_{0}(E_{\mathrm{b}}-\Delta)\right]^{-1/2}  ,\nonumber \\
\lambda_{\mathrm{c,a}} & = &
	\left[R_{0}(E_{\mathrm{t}})R_{0}(E_{\mathrm{t}}+\Delta)\right]^{-1/2}  .
\label{eq:LC_imp}
\end{eqnarray}
Here $E_{\mathrm{b}}$ and $E_{\mathrm{t}}$ are lower
and upper bare conduction
band edges, respectively, and we have taken into account that the quantities
$\mathrm{Im}R_{0}(E_{\mathrm{b}})$,
$\mathrm{Im}R_{0}(E_{\mathrm{b}}-\Delta)$,
$\mathrm{Im}R_{0}(E_{\mathrm{t}})$ and
$\mathrm{Im}R_{0}(E_{\mathrm{t}}+\Delta)$
are equal to zero.
Provided that the bare DOS has a property
\begin{equation}
N(E_{\mathrm{b}}+D+\epsilon)=N(E_{\mathrm{t}}-D-\epsilon),
\label{eq:BB_cond}
\end{equation}
(e.g, bare DOS is symmetric) the following
equalities hold 
\begin{eqnarray}
\mathrm{Re}R_{0}(E_{\mathrm{b}}-\Delta)
& = &
	-\mathrm{Re}R_{0}(E_{\mathrm{t}}+\Delta), \nonumber \\
\mathrm{Re}R_{0}(E_{\mathrm{b}})
& = &
	-\mathrm{Re}R_{0}(E_{\mathrm{t}}),
\label{eq:BB_cond1}
\end{eqnarray}
we have
\begin{equation}
\lambda_{\mathrm{c,a}}=\lambda_{\mathrm{c,b}},
\end{equation}
and hence in this case bonding and anti-bonding states
appear in the spectrum simultaneously.
For semielliptical bare DOS (\ref{eq:SEDOS}) we put
\begin{equation}
E_{\mathrm{b}}=-E_{\mathrm{t}} =D
\label{eq:SE_par}
\end{equation}
to obtain
\begin{equation}
\lambda_{\mathrm{c,a}}=\lambda_{\mathrm{c,b}} =
	\frac{D}{2} \left[\left(1+\frac{\Delta}{D}\right)
	+\sqrt{\left(1+\frac{\Delta}{D}\right)^{2}-1}\right]^{1/2} .
\end{equation}

In the dynamical  ATA (case (b)) and for $0<\Delta<D$,
we derive
\begin{equation}
\lambda_{\mathrm{c,b}}
        =  \sqrt{-\frac{2D-\Delta}{R_{0}(E_{\mathrm{b}})}}  ,
\quad
\lambda_{\mathrm{c,a}}
        = \sqrt{\frac{2D+\Delta}{R_{0}(E_{\mathrm{t}})}}
\label{eq:LC_DATA}
\end{equation}
where the conditions $\mathrm{Im}R_{0}(E_{\mathrm{b}})=0$,
$\mathrm{Im}R_{0}(E_{\mathrm{t}})=0$,
$R_{0}(E_{\mathrm{b}}) < 0$ and
$R_{0}(E_{\mathrm{t}}) > 0$ have been used. Here
$\lambda_{\mathrm{c,b}}$ and $\lambda_{\mathrm{c,a}}$ are critical
couplings for the situation where bonding and anti-bonding state
subbands have been already fully formed (e.g., the subbands are
decoupled from a resonant state subband). Provided that the
condition (\ref{eq:BB_cond1}) takes place, one has
\begin{equation}
\lambda_{\mathrm{c,b}} \le \lambda_{\mathrm{c,a}}
\end{equation}
and a bonding state subband appears in the spectrum
at lower value of $\lambda$ being compared with that
for anti-bonding state subband formation.
For semielliptical bare  DOS (\ref{eq:SEDOS},\ref{eq:SE_par})
we obtain
\begin{equation}
\lambda_{\mathrm{c,b}} =D\sqrt{1-\frac{\Delta}{2D}} ,
\quad
\lambda_{\mathrm{c,a}} = D\sqrt{1+\frac{\Delta}{2D}} .
\end{equation}

For small $\Delta$ (Fig.\ref{fig:1})
the case (c) differs from the case (b)
by a more smooth DOS behavior, which is a result of the full
self-consistency in this approach. For large $\lambda$ we obtain
in DCPA a quite rich structure owing to intrinsic impurity
dynamics at $\Delta \ne 0$, along with the standard band
splitting. One can see from  Figs.\ref{fig:1},\ref{fig:2}
that DCPA describes the
formation of the energy gap in the band centre rather adequately,
whereas  dynamical ATA yields resonance states in the ``pseudogap"
even at large $\lambda$. From analogy with disordered binary
alloys one can assume that such a behavior is owing to lack of
self-consistency in dynamical ATA (cf. the consideration of
Hubbard-III-like approximation in strongly correlated systems
\cite{Anokhin:1991:TMTPP}). At the same time, dynamical ATA gives
a reasonable picture of bonding and anti-bonding states formation.

The  non-zero values of the critical values
(\ref{eq:LC_imp},\ref{eq:LC_DATA}) are connected with the
square-root energy behavior of the bare DOS at the band edges. One
can expect that in the case of a rectangle-like bare DOS (e.g.,
two-dimensional case which may be also qualitatively described by
DMFT) the bonding and anti-bonding states will occur at
arbitrarily small $\lambda$. A similar situation near the Fermi
level is expected in the case of finite band filling in the
rigid-band approximation.

The effective mass renormalization in our model is due to band
narrowing. We obtain
\begin{equation}
\frac{m^{*}}{m} = 1 -
    \left. \frac{
    			\partial \mbox{Re}\Sigma^{\mathrm{loc}}(E)
		}{
			\partial E
		} \right|_{E=E^{*}} \le 2 .
\end{equation}
 Thus in our two-level situation the bandwidth renormalization
 does not exceed the value of two, and at large $\lambda$ only
 two subbands survive in the spectrum. At the same time,
 in the standard problem of small
 phonon  polaron (which was also considered in DMFT
 \cite{Ciuchi:1997:DMFTSP} )
increasing electron-phonon interaction results in many-fold
splitting of electron band.

To conclude, numerical results even in the simplest particular
case of our model demonstrate a rather rich and non-trivial
spectrum picture. Investigations of the general model
(\ref{eq:H0},\ref{eq:H_k})
seem to be of interest.

The work was supported in part by Grant No. 4640.2006.2
(Support of Scientific School) from the
Russian Basic Research Foundation.


\begin{thebibliography}{00}
\bibitem{Mahan:2000}
G.~D.~Mahan, {\em Many-particle physics.\/},
New York: Kluwer Academic, Plenum Publishers, 2000.


\bibitem{Maruyama:1989:EXAFS_HTC}
H. Maruyama, T. Ishii, N. Bamba  {\em et al.\/},
Physica C {\bf 160}, (1989) 524.

\bibitem{Mustre:1990:EXAFS_HTC}
J. Mustre de Leon, S. D. Conradson, I. Baltisic and A. R. Bishop,
Phys. Rev. Lett. {\bf 65}, (1990) 1675.

\bibitem{Mustre:1991:EXAFS_HTC}
J. Mustre de Leon, S. D. Conradson, I. Baltisic and A. R. Bishop,
Phys. Rev. B {\bf 44}, (1991) 2422.

\bibitem{Irkhin:1991:PKLSA}
V. Yu. Irkhin, M. I. Katsnelson and A. V. Trefilov,
Europhys. Lett., {\bf 15} (6), (1991) 649.

\bibitem{Irkhin:1994:IKT}
V. Yu. Irkhin, M. I. Katsnelson, and A. V. Trefilov,
Zh. Eksp. Theor. Fiz. {\bf 105}, (1994) 1733
[ Sov. Phys. JETP {\bf 78}, (1994) 936].


\bibitem{Kondo:1976:MG}
J. Kondo,
Physica 84 {\bf B}, (1976) 207.

\bibitem{Zawadovsky:1980:MG}
A. Zawadovsky,
Phys. Rev. Lett. {\bf 45}, (1980) 211.

\bibitem{Keiter:1980:MG}
H. Keiter and G. Morandi,
Phys. Rev. B {\bf 22}, (1980) 5004.

\bibitem{Keiter:1984:MG}
H. Keiter and G. Morandi,
Phys. Rep. {\bf 109}, (1984) 227.

\bibitem{Fulde:1985:CEF}
P. Fulde and M. Loewenhaupt,
Adv. Phys. {\bf 34}, (1985) 589;
V. Yu. Irkhin and M. I. Katsnelson,
Z. Phys. B {\bf 70}, (1988) 371.

\bibitem{Leggett:1987:TLSD}
A. J. Leggett, S. Chakravarty, A. T. Dorsey,
M. P. A. Fisher and W. Zwerger,
Rev. Mod. Phys. {\bf 59}, (1987) 1.

\bibitem{Kakehashi:1992:DCPA1}
Y. Kakehashi,
Phys. Rev. B {\bf 45}, (1992) 7196.

\bibitem{Kakehashi:1992:DCPA2}
Y. Kakehashi,
J. Magn. Magn. Mater. {\bf 104-107}, (1992) 677.

\bibitem{Kakehashi:2002:DCPA3}
Y. Kakehashi,
Phys. Rev. B {\bf 65}, (2002) 184420.

\bibitem{Kakehashi:2004:CPAP}
Y. Kakehashi and P. Fulde,
Phys. Rev. B {\bf 69}, (2004) 045101.

\bibitem{Georges:1996:DMFT}
A.Georges, G. Kotliar, W. Krauth and M. J. Rozenberg, 
Rev. Mod. Phys. {\bf 68}, (1996) 13.

\bibitem{Kakehashi:2002:COMP}
Y. Kakehashi,
Phys. Rev. B {\bf 66}, (2002) 104428.

\bibitem{Falikov:1969:SMSMT}
L. M. Falikov and J. C. Kimbal,
Phys. Rev. Lett {\bf 22}, (1969) 997.

\bibitem{Holstein:1959:MOD1}
T. Holstein,
Ann. Phys. (Leipzig) {\bf 8}, (1959) 325.

\bibitem{Holstein:1959:MOD2}
T. Holstein,
Ann. Phys. (Leipzig) {\bf 8}, (1959) 343.


\bibitem{Cini:1988:ESEPRM}
M. Cini and A. D'Andrea,
J. Phys. C: Solid State Phys. {\bf 21}, (1988) 193.

\bibitem{Eliot:1974:EKL}
R. J. Eliot, J. A. Krumhansl and P. L. Leath,
Rev. Mod. Phys. {\bf 46}, (1974) 465.

\bibitem{Anokhin:1991:TMTPP}
A. O. Anokhin, V. Yu. Irkhin and M. I. Katsnelson,
J. Phys.: Condens. Matter {\bf 3}, (1991) 1475.

\bibitem{Ciuchi:1997:DMFTSP}
S. Ciuchi, F. de Pasquale and S. Fratini, D. Feinberg,
Phys. Rev. B {\bf 56}, (1997)  4494.

\end{thebibliography}
\end{document}